\documentclass[aps,pra,superscriptaddress,10pt,floatfix,twocolumn,amsmath,amssymb]{revtex4-2}
\usepackage[utf8]{inputenc}
\usepackage{graphicx}
\usepackage[dvipsnames]{xcolor}
\usepackage[top=2cm, bottom=2cm, left=1.5cm, right=1.5cm]{geometry}
\usepackage{hyperref}
\usepackage{siunitx} 
\usepackage{todonotes} 
\usepackage[capitalise]{cleveref} 
\usepackage[normalem]{ulem} 
\usepackage{braket}
\usepackage{titlesec}
    \titleformat{\section}{\bfseries\centering\MakeUppercase}{}{}{}
    \titlespacing*{\subsection}{0em}{32pt}{*1}{}
    \titleformat{\subsection}{\bfseries}{\thesubsection.}{1em}{}
    \titlespacing*{\subsubsection}{0em}{24pt}{*1}{}
    \titleformat{\subsubsection}{\itshape}{\thesubsection.\thesubsubsection}{1em}{}

\DeclareMathOperator{\Tr}{Tr}

\begin{document}

\title{Non-Gaussian mechanical motion via single and multi-phonon subtraction \protect\\ from a thermal state}

\author{G.~Enzian$^\dagger$}
\affiliation{QOLS, Blackett Laboratory, Imperial College London, London SW7 2BW, United Kingdom}
\affiliation{Clarendon Laboratory, Department of Physics, University of Oxford, Oxford OX1 3PU, United Kingdom}
\affiliation{Niels Bohr Institute, University of Copenhagen, Copenhagen 2100, Denmark}

\author{L.~Freisem$^\dagger$}
\affiliation{QOLS, Blackett Laboratory, Imperial College London, London SW7 2BW, United Kingdom}
\affiliation{Clarendon Laboratory, Department of Physics, University of Oxford, Oxford OX1 3PU, United Kingdom}

\author{J.~J.~Price$^\dagger$}
\affiliation{QOLS, Blackett Laboratory, Imperial College London, London SW7 2BW, United Kingdom} 
\affiliation{Clarendon Laboratory, Department of Physics, University of Oxford, Oxford OX1 3PU, United Kingdom}

\author{A.~\O.~Svela$^\dagger$}
\affiliation{QOLS, Blackett Laboratory, Imperial College London, London SW7 2BW, United Kingdom}
\affiliation{Clarendon Laboratory, Department of Physics, University of Oxford, Oxford OX1 3PU, United Kingdom}
\affiliation{Max Planck Institute for the Science of Light, Staudtsta{\ss}e 2, 91058 Erlangen, Germany}

\author{J.~Clarke}
\affiliation{QOLS, Blackett Laboratory, Imperial College London, London SW7 2BW, United Kingdom}

\author{B.~Shajilal}
\author{J.~Janousek}
\author{B.~C.~Buchler}
\author{P.~K.~Lam}
\affiliation{Centre for Quantum Computation and Communication Technology, Research School of Physics and Engineering, Australian National University, Canberra 2601, Australia}

\author{M.~R.~Vanner}\email{www.qmeas.net (m.vanner@imperial.ac.uk)}
\affiliation{QOLS, Blackett Laboratory, Imperial College London, London SW7 2BW, United Kingdom}
\affiliation{Clarendon Laboratory, Department of Physics, University of Oxford, Oxford OX1 3PU, United Kingdom}

\begin{abstract}
Quantum optical measurement techniques offer a rich avenue for quantum control of mechanical oscillators via cavity optomechanics. In particular, a powerful yet little explored combination utilizes optical measurements to perform heralded non-Gaussian mechanical state preparation followed by tomography to determine the mechanical phase-space distribution. Here, we experimentally perform heralded single- and multi-phonon subtraction via photon counting to a laser-cooled mechanical thermal state with a Brillouin optomechanical system at room temperature, and use optical heterodyne detection to measure the $s$-parameterized Wigner distribution of the non-Gaussian mechanical states generated. The techniques developed here advance the state-of-the-art for optics-based tomography of mechanical states and will be useful for a broad range of applied and fundamental studies that utilize mechanical quantum-state engineering and tomography.
\end{abstract}

\maketitle

\textit{Introduction.}---A key current goal in cavity quantum optomechanics is to generate and fully characterize non-Gaussian states of mechanical motion that exhibit non-classical behavior. Pursuing this line of research will facilitate the development of mechanical-oscillator-based quantum technology such as quantum memories exploiting the long coherence times available~\cite{Galliou2013, Renninger2018, MacCabe2020}, coherent transducers~\cite{Higginbotham2018, Mirhosseini2020}, and sensors~\cite{Kim2016, Monteiro2017, Carney2021}. Additionally, such state generation and characterization capabilities will help explore fundamental physics including the quantum-to-classical transition~\cite{BJK99, Marshall2003, Bassi2013} and even the interface between quantum mechanics and gravity~\cite{Pikovski2012, Bose2017, Marletto2017}.

Throughout quantum optics, non-Gaussian state preparation of a bosonic mode followed by phase-space characterization has been performed with a wide spectrum of different platforms. For trapped ions, a single-phonon Fock state of motion of was prepared and reconstructed~\cite{Leibfried1996}, and multi-component superposition states are now being studied~\cite{Fluhmann2019}. In optics, heralded single-photon addition or subtraction followed by homodyne tomography has been widely utilized, with prominent examples including: Wigner tomography of a heralded single-photon state~\cite{Lvovsky2001}, single- and multi-photon subtraction to squeezed states to generate superposition states~\cite{Ourjoumtsev2006, Neergaard2006, Gerrits2010}, photon-addition and subtraction to optical thermal states~\cite{Zavatta2007, Parigi2007, Bogdanov2017}. Other notable examples of non-Gaussian quantum states with other physical systems include: studying the decoherence of a superposition state of a microwave field inside a cavity~\cite{Deleglise2008}, generating non-Gaussian states of atomic-spin ensembles~\cite{McConnell2015}, creating arbitrary quantum states in a microwave superconducting circuit~\cite{Hofheinz2009}, and creating non-classical states of acoustic waves coupled to superconducting qubits~\cite{Chu2018, Satzinger2018}.

Within optomechanics excellent progress utilizing single-photon detection has been made including generating non-classical states of high-frequency vibrations in diamond crystals~\cite{Lee2012, Fisher2017, Velez2019} and photonic-crystal structures~\cite{Riedinger2016}, second-order-coherence measurements of mechanical modes~\cite{Cohen2015, Galinskiy2020}, the generation of mechanical interference fringes~\cite{Ringbauer2018}, and single-phonon addition or subtraction to a thermal state resulting in a doubling of the mean occupation~\cite{Enzian2021}. There is also significant progress towards developing the experimental tools needed for mechanical phase-space tomography or reconstruction~\cite{Vanner2013, Muhonen2019, Suchoi2015, Rashid2017, Ringbauer2018}; however, all of these experiments have insufficient sensitivity to resolve features below the mechanical zero-point motion, and phase-space characterization~\cite{Vanner2015} of a mechanical quantum state remains outstanding within in optomechanics. One route to achieve this goal in the resolved-sideband regime is to perform single-phonon addition or subtraction for quantum-state preparation and then utilize a red-sideband drive and optical state tomography with a balanced detector, such as homodyne or heterodyne detection.

In this Letter, we describe an experimental study that observes non-Gaussian phase-space distributions generated by single- and multi-phonon subtraction to a thermal state of a mechanical oscillator. These operations are heralded by single- and multi-photon detection events following an optomechanical interaction to a laser-cooled state from room temperature. We utilize quantum-noise-limited heterodyne detection to characterize the mechanical states prepared, and advance the state-of-the-art for optics-based mechanical state tomography by more than an order of magnitude in terms of overall efficiency and added noise. We observe that the initial thermal state is transformed by these operations from a Gaussian in phase space into a ring shape with a diameter that increases with the number of phonons subtracted. Building on established results in quantum optics and recent work demonstrating that the mechanical mean occupation doubles for single-phonon addition and subtraction~\cite{Enzian2021}, here we additionally make the first observation that the mean occupation triples for two-phonon subtraction. This work expands the toolkit for optical control and readout of mechanical states, and can be applied to experiments to exploit and characterize the non-Gaussian and non-classical properties these operations generate.

\begin{figure}[t]
    \centering
    \includegraphics[width=75mm]{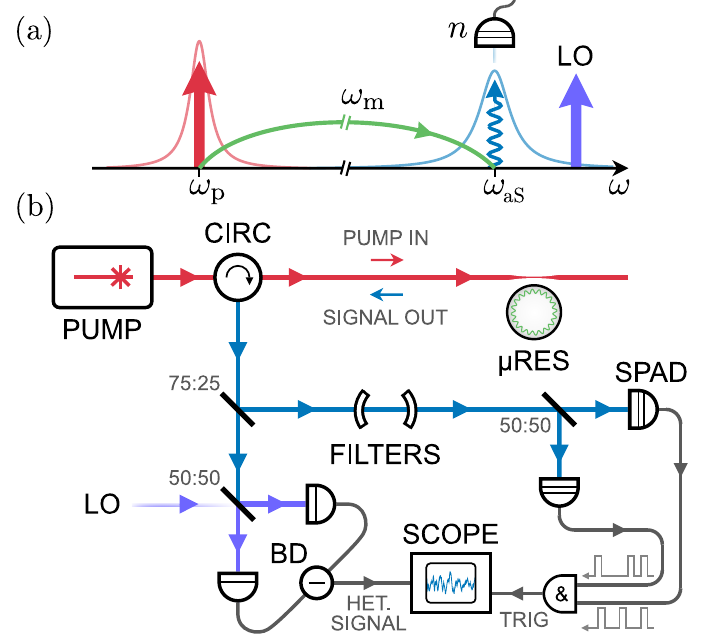}
    \caption{\small Multi-phonon subtraction and tomography scheme and experimental schematic. 
    (a) Optical pumping and heralding scheme. A pair of optical resonances spaced by the mechanical angular frequency $\omega_\textrm{m}$ are used to resonantly enhance the optomechanical interaction. A pump field drives a mode at $\omega_\textrm{p}$ creating an anti-Stokes signal at $\omega_\textrm{s}$. An $n$-photon detection scheme is then used to herald $n$-phonon subtraction to the mechanical motion.
    (b) Experimental schematic. A tuneable pump laser (1550~nm) drives an optical microresonator and the backscattered anti-Stokes signal is separated from the pump with an optical circulator (CIRC). The signal is subsequently split and detected via single photon avalanche detectors (SPADs) to herald a single- or two-phonon subtraction operation. To characterize the mechanical state prepared, heterodyne detection is performed (BD: balanced detector, LO: local oscillator), and the signal recorded on an oscilloscope triggered by SPAD detection events. The two-phonon subtraction case is shown here that uses a two-photon-coincidence measurement.
    }
    \label{Fig:SchemeSetup}
\end{figure}

\textit{Multi-phonon subtraction scheme.}---To perform $n$-phonon subtraction, we use a pair of optical cavity modes that are approximately spaced by the mechanical frequency to resonantly enhance the optomechanical interaction (cf.~Fig.~\ref{Fig:SchemeSetup}(a)). Each optical mode has a linewidth much smaller than the mechanical frequency allowing operation deeply within the resolved-sideband regime. By optically driving the lower-frequency mode of the pair, the anti-Stokes scattering process is selected and a signal field in the higher-frequency cavity mode is generated. This scattering process is described by a light-mechanics beamsplitter interaction with Hamiltonian $H/\hbar = G(a^\dagger b + a b^\dagger)$, where $G$ is the linearized optomechanical coupling rate, $a$ is the field operator of the optical signal, and $b$ is the mechanical annihilation operator. Then, detecting $n$ photons heralds an $n$-phonon subtraction to the mechanical state. For an initial mechanical thermal state $\rho_{\bar{n}}$, with mean occupation $\bar{n}$, the mechanical state following this operation may be written $\rho_{n-} \propto b^n \rho_{\bar{n}} b^{\dagger n}$. Similarly, by optically driving the upper-frequency mode, the Stokes scattering process is selected ($H/\hbar = G(a b + a ^\dagger b^\dagger)$) and detecting $n$ photons in the frequency down-shifted Stokes signal heralds $n$-phonon addition, $\rho_{n+} \propto b^{\dagger n} \rho_{\bar{n}} b^{n}$. These operations are a multi-phonon generalization to Ref.~\cite{VannerKim2013}.

When performing single- or multi-phonon subtraction or addition to a large thermal state, one may expect little change to the mechanical state. However, these operations significantly change the mean occupation and give rise to highly non-Gaussian distributions in mechanical phase space. Indeed, when applying an $n$-fold subtraction operation, the mean occupation transforms via $\bar{n} \rightarrow (n+1)\bar{n}$, and for $n$-fold addition $\bar{n} \rightarrow (n+1)\bar{n}+n$. For $\bar{n} \gg 1$, it is noted that the mean occupation \textit{doubles} for single-quanta addition or subtraction---which has been experimentally observed for thermal optical fields~\cite{Zavatta2007} and very recently for a mechanical thermal state~\cite{Enzian2021}---and the mean occupation \textit{triples} for two-quanta addition or subtraction. This significant change to the mean occupation and the non-Gaussian ring shape observed in phase space can be understood via a combination of the shift to the probability distribution of the number operator and the Bayesian inference of the non-unitary quantum-measurement process~\cite{Barnett2018}. It is also important to note that in the limit $\bar{n} \gg \sqrt{n}$ the mechanical state generated by $n$-phonon addition is approximately the same as that generated by $n$-phonon subtraction. Thus, for this work, we focus on $n$-phonon subtraction and mechanical state readout using the anti-Stokes interaction as this interaction is well suited for mechanical state readout and the optical field is a high-fidelity proxy for the mechanical state in the limit of high efficiency. See the Supplementary Material~\cite{Supp} for more mathematical details.

\textit{Experimental setup.}---In this work, $n$-phonon subtraction is implemented by driving anti-Stokes Brillouin scattering in a BaF$_2$ optical microresonator. BaF$_2$ is an attractive material for these studies as the low optical and acoustic losses enable significant enhancement of Brillouin optomechanical interaction~\cite{Lin2014, Supp}. Here, we use a pair of optical resonances (amplitude decay rates $\kappa_\mathrm{p}/2\pi = 7.1$~MHz, $\kappa_\mathrm{aS}/2\pi = 46.9$~MHz) with a separation approximately equal to the mechanical frequency, $\omega_\text{m}/2\pi = 8.16$~GHz. The mechanical amplitude decay rate and intrinsic optomechanical coupling rate are $\gamma/2\pi = 3.26(39)$~MHz and $g_0/2\pi = 296(39)$~Hz, respectively~\cite{Supp}. The experiment was performed at $300$~K, corresponding to a mean mechanical thermal occupation of $\bar{n}_\text{th} \simeq 766$, which, via the optomechanical coupling was sideband-cooled to $\bar{n} \simeq 453(52)$. An input pump power of $\sim$9~mW was used, corresponding to an intra-cavity photon number of $N_\text{cav} \simeq 1.2 \times 10^9$ and optomechanical coupling rate of $G / 2\pi \simeq 10$~MHz, such that the system is well within the weak coupling regime ($2G<\kappa_\mathrm{aS}+\gamma$). The backscattered anti-Stokes signal is coupled out of the cavity by a silica tapered fiber, with efficiency $\eta_\text{c}~\simeq~0.25$, and is then separated from the forwards-propagating pump field using an optical circulator before being split into two arms using a 75:25 beamsplitter (cf.~\cref{Fig:SchemeSetup}(b)). Thus, the anti-Stokes field serves a dual role where one arm is used for heralded mechanical state preparation via photon counting; and the other arm is used for mechanical state tomography via heterodyne detection.

\begin{figure*}[ht]
    \centering
    \includegraphics[width=\textwidth]{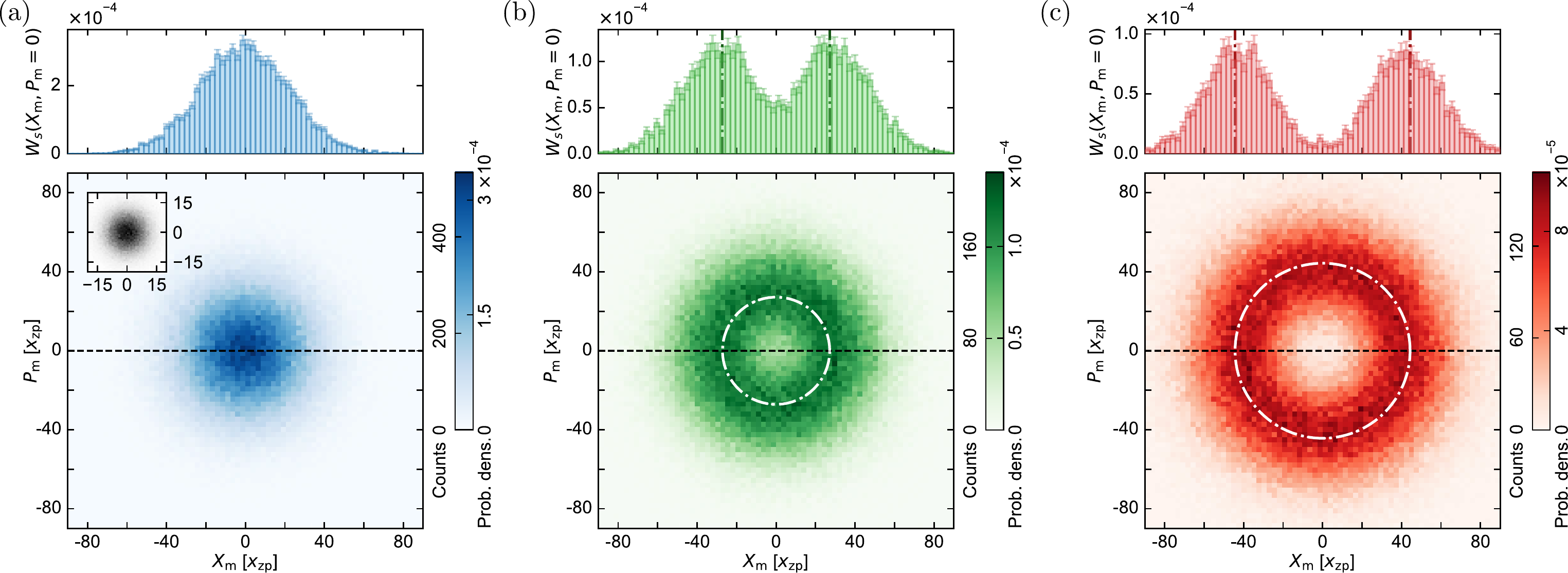}
    \vspace{-6mm}
    \caption{\small Experimental $s$-parameterized Wigner functions $W_s$ (bottom row), with slices through $P_\mathrm{m}=0$ (top row), for an (a)~initial, (b)~single-phonon subtracted and (c)~two-phonon subtracted mechanical thermal state.
    The phase-space distributions are plotted in units of the mechanical zero-point fluctuations $x_\text{zp}$ and are obtained through heterodyne detection of the optical anti-Stokes signal. The generation of non-Gaussianity from the originally Gaussian phase-space distribution is observed for single-phonon subtraction, which further grows upon two-phonon subtraction. The dash-dotted lines indicate the theoretically predicted maxima for the $W_s$ functions. The phase-space distribution of the optical vacuum contribution is shown in the inset in panel (a).
    }
    \label{Fig:PhaseSpace}
\end{figure*}

In the 25\% arm, single- and two-phonon subtraction events are heralded using two single-photon avalanche detectors (SPADs). Prior to the detectors, two fiber-based Fabry--Perot filters are used in series to filter out deleterious photons from the pump-field. As the single-photon count rate ($280(40)$~s$^{-1}$) is much greater than the dark count rate ($\sim$~1~s$^{-1}$), the phonon-subtraction events are heralded with high fidelity. Note that as the gate window of 3.5~ns is much less than the decay time $1/(\gamma+G^2/\kappa_\mathrm{aS}) \simeq 31$~ns, the heralding of two-photon events is well approximated as simultaneous detections. See~\cite{Supp} for further details regarding these heralding rates. Also, note that any photons that are not detected here do not result in phonon subtraction but rather contribute to mechanical laser cooling.

In the 75\%~arm, balanced heterodyne detection is used to perform phase-space tomography of the mechanical states generated. The detection scheme is implemented by interfering the anti-Stokes signal and a strong local oscillator, that is frequency-detuned by $\omega_\text{het}/2\pi = 214$~MHz with respect to the signal, onto a 50:50 beamsplitter and measuring the output using a balanced photodetector. The two types of photon counting events---singles and two-photon coincidences---then trigger a high-bandwidth oscilloscope to record a time-trace of the output from the balanced detector. In order to acquire sufficient statistics for the phase-space distributions and temporal dynamics of the mechanical states for the initial, single-phonon subtracted, and two-phonon subtracted thermal states, \num{2.4e5} time traces were recorded for each case.

\textit{Mechanical state readout.}---One promising route to perform mechanical quantum state characterization is to use optical homodyne or heterodyne tomography, as utilized in quantum optics, after having performed an efficient transfer of the mechanical state to the optical field. In the absence of losses or inefficiencies, if optical homodyne tomography is performed after the state transfer, then the marginals obtained allow the mechanical Wigner function to be reconstructed. And, if a heterodyne measurement is performed, owing to the vacuum noise introduced with the simultaneous measurement of the two conjugate optical quadratures, the Husimi-$Q$ function is obtained. 

In practice, one of the most important aspects to such a measurement is the overall efficiency, as any loss and inefficiency reduces the quality of the phase-space distribution obtained. More specifically, optical losses result in the state of interest being convolved in phase space with vacuum noise, thus degrading the signal and even eliminating any non-classical features if the efficiency is poor.

A versatile way to mathematically quantify the performance of a tomography experiment is to use the $s$-parameterized Wigner function $W_s(X_\textrm{m}, P_\textrm{m})$~\cite{Leonhardt1993}, as it captures the unwanted effects of noise and inefficiency in a single parameter. For our experiment here, which is limited by efficiency rather than any additional noise, the $s$-parameter is defined as $s = (\eta-2)/\eta$, where $\eta$ is the overall measurement efficiency of the mechanical state including the mechanics-light transduction efficiency. From the expression for $s$, we can see that for $\eta = 1$ we have $s = -1$, which corresponds to the $Q$ function, and for $\eta < 1$ we have $s < -1$ corresponding to a distribution that is smoother than the $Q$ function. Experimentally determining $W_s$ for mechanical states fully characterizes the state allowing any statistic or measurement probability to be determined, and will aid in mechanical quantum state engineering applications. For $n$-phonon subtraction, the $s$-parametrized Wigner function can be written as a two-dimensional convolution between the $P$ function of the subtracted state and a Gaussian, $W_s(X_\textrm{m}, P_\textrm{m}) = \left(P_{n-} * G_s\right)(X_\textrm{m}, P_\textrm{m})$, where $P_{n-} = (2^{-n}\bar{n}^{-n-1}/\pi n!) \left(X_\textrm{m}^2+P_\textrm{m}^2\right)^n e^{ -(X_\textrm{m}^2+P_\textrm{m}^2)/2\bar{n}}$ 
and $G_{s}=e^{-(X_\textrm{m}^2+P_\textrm{m}^2)/(1-s)}/\pi (1-s)$~\cite{Supp}.

In this experiment, via the anti-Stokes light-mechanics beam-splitter interaction, the scattered optical signal acts as a proxy for the acoustic mode and is used to characterize the mechanical state generated at the time of the herald event. From the time-domain heterodyne signal we extract the two orthogonal quadrature components by demodulating the signal at the heterodyne frequency in post-processing for each herald event. The quadrature signals are then used for two types of analysis. Firstly, we compute the variance of the ensemble of measurements for each time about the herald event, and secondly, a two-dimensional histogram of the mechanical phase-space distribution $W_s$ at the heralding time, is obtained.

\textit{Results and discussion.}---It is first instructive to discuss 
the overall efficiency of the mechanical state tomography $\eta$. Since we are performing a heterodyne measurement, the variance of the signal, in the steady-state, away from a heralding event is $\sigma^2 = \eta \bar{n}_\mathrm{th} + 1$. For this experiment, we determine an overall efficiency $\eta = 0.91\%$ via independent measurement of the optical vacuum, and knowledge of $\bar{n}_\mathrm{th}$. This efficiency yields an $s$-parameter of $s = -219$, corresponding to a 15 times improvement to the forefront of optics-based mechanical tomography, Ref.~\cite{Muhonen2019}, which utilized fast pulsed measurements outside the resolved sideband regime. Knowing $\eta$ also allows one to work with units of the mechanical zero-point fluctuations $x_\textrm{zp}$, rather than units of the optical vacuum, by scaling the heterodyne signals accordingly. Using this efficiency, the experimentally determined mechanical phase-space distributions $W_s$ are plotted in Fig.~\ref{Fig:PhaseSpace} for the initial thermal state, the single-phonon subtracted state, and the two-phonon subtracted state. Note the highly non-Gaussian ring shape, which has an increasing radius from one- to two-phonon subtraction. Theoretical predictions computed in this work for the radii of the phase-space rings~\cite{Supp} are indicated by the dash-dotted lines in Fig.~\ref{Fig:PhaseSpace}.

\begin{figure*}[ht]
    \centering
    \includegraphics[width=0.97\textwidth]{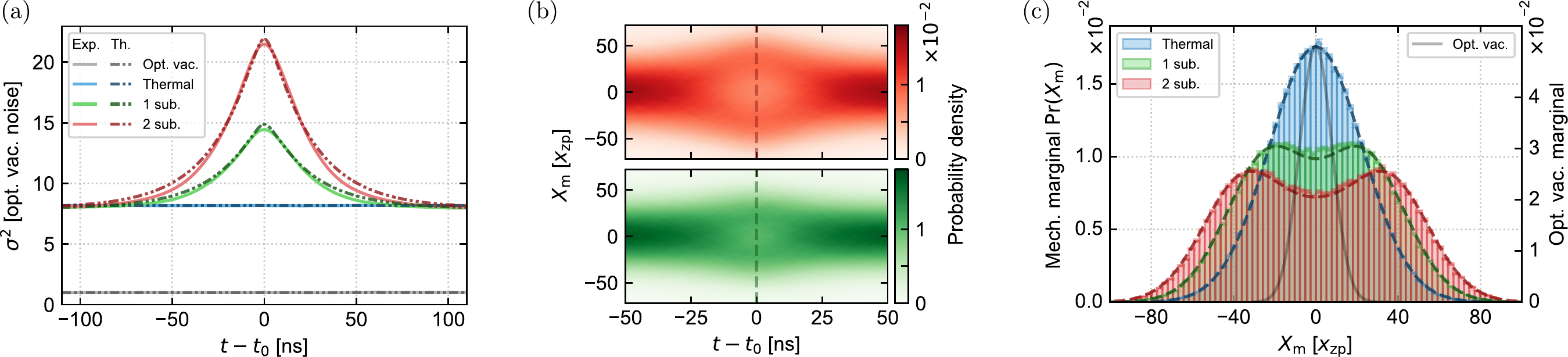}
    \vspace{-3mm}
    \caption{\small
    Dynamics and non-Gaussian distributions of the heralded mechanical states.
    (a) Time evolution of the heterodyne variance in units of optical vacuum noise about the heralding event for single- and two-phonon subtracted thermal states, plotted in green and red, respectively. The optical vacuum is plotted in gray and the variance of the initial thermal state is plotted in blue. Experimentally obtained variances are shown as solid lines, and the predictions of our theoretical model are shown as dotted-dashed lines. The Poissonian experimental relative uncertainty is of order \num{e-3} and is not visible on this scale. The discrepancy between theory and experiment is attributed to filtering in post-processing and dark counts from the SPADs. At the time of the heralding event, the ratio of the heterodyne variance to the optical vacuum noise increases by a factor of 1.94 and 2.94 relative to the initial thermal state for single- and two-phonon subtraction, respectively.
    (b) Marginal distributions for the $X_\mathrm{m}$-quadrature of the mechanical oscillator as a function of time. (c) Mechanical quadrature probability distributions at the time of the heralding event, $t=t_0$, for the initial (blue), single-phonon subtracted (green), and two-phonon subtracted (red) thermal mechanical states.
    }
    \label{Fig:MarginalsVariance}
\end{figure*}

Figure~\ref{Fig:MarginalsVariance}(a) shows the heterodyne signal variance $\sigma^2$, normalised in units of the optical vacuum noise, as a function of time about the herald event. For the single- and two-phonon subtraction cases, it is observed that the variance increases at the time of the herald event by a factor of 1.94 and 2.94, respectively, compared to the mean variance of $7.96$. This is in close agreement with the variance `doubling' and `tripling' from the theoretical predictions~\cite{Supp} shown, for which only the heralding time $t_0$ is a free fitting parameter. We attribute the small difference between our experimental observations and the theoretical prediction to be due to the filtering in the quadrature demodulation, the small level of dark counts in the SPAD detectors, and the optical filtering performed in the heralding arm. From this variance, it is seen that the contribution from optical vacuum is $14\%$, or equivalently, the overall measurement efficiency yielded a total added noise of $|s|/2 = 110$ mechanical quanta.

In Fig.~\ref{Fig:MarginalsVariance}(b), we have plotted the marginals $\mathrm{Pr}(X_\mathrm{m}) = \int d P_\mathrm{m} W_s(X_\mathrm{m},P_\mathrm{m})$ as a function of time about the herald event. This plot illustrates how the state transforms by the single- and two-phonon subtraction operations from an initial Gaussian state to a non-Gaussian state that has a bimodal quadrature probability distribution and then returns to thermal equilibrium. In Fig.~\ref{Fig:MarginalsVariance}(c), the mechanical quadrature probability distributions $\textrm{Pr}(X_\textrm{m})$ at the time of the herald event are plotted. At this time, the non-Gaussianity generated is most significant and the initial distribution can be compared with the bi-modal distributions generated via single- and two-phonon subtraction together with the theoretical prediction~\cite{Supp} overlaid.

\vspace{1em}

\textit{Conclusions and outlook.}---Utilizing both photon counting and optical heterodyne measurements, we report the first experimental generation and phase-space tomography of non-Gaussian states of mechanical motion via single- and two-phonon subtraction to a laser-cooled thermal state. In achieving this milestone this work advances optics-based mechanical tomography by more than an order of magnitude in the $s$-parameter. These advancements make key steps towards mechanical phase-space tomography of nonclassical mechanical states, which remains outstanding within optomechanics. Furthermore, the techniques developed here can be utilized for a wide range of mechanical quantum-state engineering applications taking advantage of single- and multiple-phonon addition and subtraction operations. In particular, these operations can be applied to a mechanical squeezed state for superposition state preparation~\cite{Milburn2016}, and reservoir engineering has been discussed as a promising route to generate the squeezing in such protocols~\cite{Shomroni2020, Zhan2020}. 

For this experiment, we would like to highlight four key pathways for improvement: (i) operation at cryogenic temperature to reduce the material contributions to the mechanical decay rate~\cite{Ohno2006, Galliou2013, Renninger2018, MacCabe2020}, (ii) increasing the drive strength and utilizing the optomechanical strong coupling available in Brillouin optomechanical systems~\cite{Enzian2019}, (iii) further tapered fibre and microresonator optimization to enable better optical coupling with respect to the intrinsic cavity losses, and (iv) utilizing a Stokes interaction for mechanical state preparation, followed by an anti-Stokes interaction for the readout, to provide a route to make the readout more independent and remove the beam-splitter for the single-photon detection to improve the efficiency. Implementing these four improvements provides a promising path to achieving an overall anti-Stokes readout measurement efficiency exceeding $50\%$. Achieving this efficiency, together with performing quantum-noise-limited homodyne detection, yields an $s$-parameter of $s>-1$, which is required to observe negativity of a quantum phase-space-distribution---a key signature of non-classicality and a powerful resource for quantum-enhanced technologies. Additionally, achieving a higher-efficiency anti-Stokes interaction with the paths above provides a means to implement a quantum memory device that can efficiently write and read quantum states to and from the acoustic mode.

\textit{Acknowledgements.}---We acknowledge useful discussions with P.~Del’Haye, M.~S.~Kim, J.~Nunn, N.~Moroney, A.~Rauschenbeutel, P.~Schneeweiss, J.~Silver, and S.~Zhang. This project was supported by the Engineering and Physical Sciences Research Council (EP/T031271/1, EP/P510257/1), UK Research and Innovation (MR/S032924/1), the Royal Society, the Aker Scholarship, EU Horizon 2020 Program (847523 ‘INTERACTIONS’), and the Australian Research Council (CE170100012, FL150100019). ($^\dagger$) G.E., L.F., J.J.P, and A.\O.S. contributed equally to this work and are listed alphabetically.

\textit{Note added.}---During the preparation of this manuscript we became aware of related experimental work also observing mechanical non-Gaussianity~\cite{Patel2021}.


\onecolumngrid
\clearpage
\newgeometry{left=3.5cm,right=3.5cm,top=3cm,bottom=3cm}
\setcounter{equation}{0}
\def\theequation{S\arabic{equation}}
\pagenumbering{roman}

\section*{Supplementary Material}
\begin{centering}
\textbf{
Non-Gaussian mechanical motion via single and multi-phonon subtraction \\ from a thermal state \\ 
}
\end{centering}
\vspace{10pt}

\begin{centering}
G.~Enzian$^{\dagger},^{1,2,3}\ $ 
L.~Freisem$^{\dagger},^{1,2}\ $ 
J.~J.~Price$^{\dagger},^{1,2}\ $ 
A.~\O.~Svela$^{\dagger},^{1,2,4}\ $ 
J.~Clarke,$^1$\\
B.~Shajilal,$^5\ $ 
J.~Janousek,$^5\ $ 
B.~C.~Buchler,$^5\ $ 
P.~K.~Lam,$^5\ $ 
M.~R.~Vanner$^{1,2,\ast}$\\
\end{centering}

\vspace{16pt}

\addcontentsline{toc}{section}{Supplementary Material}

\begin{centering}
\textit{\small
$^1$QOLS, Blackett Laboratory, Imperial College London, London SW7 2BW, UK\\
$^2$Clarendon Laboratory, Department of Physics, University of Oxford, Oxford OX1 3PU, UK\\
$^3$Niels Bohr Institute, University of Copenhagen, Copenhagen 2100, Denmark\\
$^4$Max Planck Institute for the Science of Light, Staudtsta{\ss}e 2, 91058 Erlangen, Germany\\
$^5$Centre for Quantum Computation and Communication Technology, Research School of Physics and Engineering, Australian National University, Canberra 2601, Australia\\
}
\end{centering}

\subsection{Multi-phonon subtraction and addition to a thermal state}
We consider an initial sideband-cooled thermal state $\rho_{\bar{n}}$ of a mechanical oscillator with mean occupation number $\bar{n}$
\begin{eqnarray}
\rho_{\bar{n}}&=&\dfrac{1}{1+\bar{n}}\sum_{m=0}^{\infty}\left(\dfrac{\bar{n}}{1+\bar{n}}\right)^m|{m}\rangle\langle{m}|\\
&=&(1-x)\sum_{m=0}^{\infty}x^m|{m}\rangle\langle{m}|.
\end{eqnarray}
Here, $x=\bar{n}/(1+\bar{n})$. An $n$-phonon subtraction operation to $\rho_{\bar{n}}$ creates the state
\begin{eqnarray}\label{rhosubeq}
\rho_{n-}&=&\dfrac{b^n \rho_{\bar{n}} b^{\dagger n}}{\Tr\left(b^n \rho_{\bar{n}} b^{\dagger n}\right)},
\end{eqnarray}
while an $n$-phonon addition operation yields
\begin{eqnarray}
\rho_{n+}&=&\dfrac{b^{\dagger n} \rho_{\bar{n}} b^{n}}{\Tr\left(b^{\dagger n} \rho_{\bar{n}} b^{n}\right)}.
\end{eqnarray}
The mean phonon number of the $n$-phonon-subtracted state $\rho_{n-}$ is given by 
\begin{eqnarray}\label{nsubeq}
\langle b^{\dagger}b\rangle_{n-}&=&\dfrac{\Tr\left(b^{n+1} \rho_{\bar{n}} b^{\dagger n+1}\right)}{\Tr\left(b^n \rho_{\bar{n}} b^{\dagger n}\right)}.
\end{eqnarray}
Here, the trace terms $\Tr\left(b^k \rho_{\bar{n}} b^{\dagger k}\right)$ with $k\in\mathbb{N}$, may be calculated by using $b^{k}|{m}\rangle=\sqrt{m!/(m-k)!}|{m-k}\rangle$ for $m\geq k$ and $b^{k}|{m}\rangle=0$ otherwise, to arrive at
\begin{eqnarray}
\Tr\left(b^k \rho_{\bar{n}} b^{\dagger k}\right)
&=&(1-x)\sum_{m=0}^{\infty}\dfrac{m!}{(m-k)!}x^m \\
&=&\dfrac{k!}{(1-x)^k}x^k.
\end{eqnarray}
This expression for the trace may be inserted into Eqs.~\eqref{rhosubeq} and \eqref{nsubeq} to give the explicit form of the density operator $\rho_{n-}$ and the mean phonon number $\langle b^{\dagger}b\rangle_{n-}$
\begin{eqnarray}
\rho_{n-}&=&\sum_{m=0}^{\infty}p_{n-}(m)|{m}\rangle\langle{m}|,\\
p_{n-}(m)&=&(1-x)^{n+1}x^m\binom{m+n}{n},\\
\langle b^{\dagger}b\rangle_{n-}&=&(n+1)\bar{n}.
\end{eqnarray}
A similar calculation allows one to calculate the density operator and the mean phonon number of the $n$-phonon-added state $\rho_{n+}$, which are given by
\begin{eqnarray}
\rho_{n+}&=&\sum_{m=0}^{\infty}p_{n+}(m)|{m}\rangle\langle{m}|,\\
p_{n+}(m)&=&(1-x)^{n+1}x^{m-n}\binom{m}{n},\\
\langle b^{\dagger}b\rangle_{n+}&=&(n+1)\bar{n}+n.
\end{eqnarray}
Here, the binomial coefficient $\binom{m}{n}=0$ for $m<n$, such that phonon-number states $|m\rangle$ with $m<n$ of $\rho_{n+}$ are unoccupied. This property also leads to the observation that the phonon-number distribution of $\rho_{n+}$ and $\rho_{n-}$ are equal up to a shift $p_{n+}(m)=p_{n-}(m-n)$. We also note that to further generalize these results, moment-generating functions are a valuable and elegant tool for these types of statistics [S.~M.~Barnett, G.~Ferenczi, C.~R.~Gilson, and F.~C.~Speirits, Phys.~Rev.~A \textbf{98}, 013809 (2018)].

To further study the similarity of the states $\rho_{n-}$ and $\rho_{n+}$ we calculate the quantum state fidelity between them. By using $[\rho_{n-},\rho_{n+}]=0$, one has that the fidelity is $F(\rho_{n-},\rho_{n+})=\sum_{m}\sqrt{p_{n-}(m)p_{n+}(m)}$, which is strictly less than one. Utilizing $\binom{m+n}{n}<\binom{m}{n}$, we then derive a lower bound for the fidelity
\begin{eqnarray}
\left(\dfrac{\bar{n}}{1+\bar{n}}\right)^{n/2}<F(\rho_{n-},\rho_{n+})<1.
\end{eqnarray}
For finite $n$, this lower bound approaches one from below as $\bar{n}$ increases, and hence we conclude that the two states are approximately the same in the high $\bar{n}$ limit. Moreover, to arrive at a stringent condition on $\bar{n}$ and $n$ for the states $\rho_{n-}$ and $\rho_{n+}$ to be the same, we demand that the separation of the means $\langle b^{\dagger}b\rangle_{n+}-\langle b^{\dagger}b\rangle_{n-}$ is much less than the variance in phonon number $\mathrm{Var}\left(\rho_{n+},b^{\dagger}b\right)=\mathrm{Var}\left(\rho_{n-},b^{\dagger}b\right)$. This gives $\bar{n}\gg\left(-(1+n)+\sqrt{4n^3+5n^2+2n+1}\right)/\left(2(1+n)\right)$, where the right hand side of this inequality is an increasing function of $n$. Hence, at high values of $n$, we arrive at the condition for the mechanical state generated by an $n$-phonon subtraction to a thermal state to be the same as that generated by an $n$-phonon addition operation, namely $\bar{n}\gg\sqrt{n}$.

\subsection{Marginal distributions}

\subsubsection{Position marginals}

The Glauber--Sudarshan $P$ function $P(\beta)$ of a quantum state is related to its density operator $\rho$ through 
\begin{eqnarray}
\rho&=&\int{d}^2\beta~ P(\beta)|{\beta}\rangle\langle{\beta}|.
\end{eqnarray}
For example, the $P$ function of $\rho_{\bar{n}}$ is given by $P_{\bar{n}}(\beta)=\mathrm{e}^{-|\beta|^2/\bar{n}}/\pi\bar{n}$.
Using Eq.~\eqref{rhosubeq}, one finds that the $P$ function of $\rho_{n-}$ is
\begin{equation}
    \label{eq:Pfuncnsub}
    P_{n-}(\beta) = \dfrac{1}{n!\bar{n}^n}|\beta|^{2n}P_{\bar{n}}(\beta).
\end{equation}
Note that $P_{n-}(\beta)$ only depends on the magnitude of $\beta$, which demonstrates the rotational symmetry of the state in phase space. 

We then calculate the position marginal of $\rho_{n-}$ to be
\begin{eqnarray}
&&\mathrm{pr}_{n-}(X_\textrm{m})=\int{d}^2\beta~ P_{n-}(\beta)|\langle{X_\textrm{m}}|{\beta}\rangle|^2\label{probPfunceq}\\
&=&\dfrac{\exp\left({-\frac{X_\textrm{m}^2}{1+2\bar{n}}}\right)}{n!\pi^{\frac{3}{2}}\sqrt{1+2\bar{n}}}\sum_{k=0}^{n}\sum_{l=0}^{k}\binom{n}{k}\binom{2k}{2l}
\Gamma\left[n-k+\frac{1}{2}\right]\Gamma\left[l+\frac{1}{2}\right]
X_\textrm{m}^{2(k-l)}\dfrac{(2\bar{n})^{k-l}}{(1+2\bar{n})^{2k-l}}.
\end{eqnarray}
Where the gamma functions are given by
\begin{eqnarray}
\Gamma\left[m+\frac{1}{2}\right]=\dfrac{(2m)!}{4^mm!}\sqrt{\pi}
\end{eqnarray}
for $m\in\mathbb{N}$.
Due to the rotational symmetry of the state, the probability marginals are invariant under the transformation $X_\textrm{m}\rightarrow X_\textrm{m}({\theta})= X_\textrm{m}\cos\theta+P_\textrm{m}\sin\theta$. 

\subsubsection{Inefficiencies in measurement}
The total measurement efficiency $\eta$ of the mechanical state includes mechanics-light transduction efficiency and optical detection efficiencies. An inefficient measurement of the mechanical state marginal $\mathrm{pr}(X_\textrm{m})$ is described by a beamsplitter model for loss
\begin{eqnarray}
\mathrm{pr}(X_\textrm{m};\eta)&=&\dfrac{1}{\sqrt{\pi(1-\eta)}}\int_{-\infty}^{+\infty}{d}{X'}~\mathrm{pr}(X')\exp\left(-\frac{\eta}{1-\eta}(X'-X_\textrm{m}/\sqrt{\eta})^2\right),\label{Bsmodelloss}
\end{eqnarray}
where $\mathrm{pr}(X_\textrm{m};\eta)$ is the marginal of the optical field to be measured~[U.~Leonhardt, \emph{Measuring the Quantum State of Light} Cambridge University Press (1997)]. Inserting $\mathrm{pr}_{n-}(X_\textrm{m})$ into Eq.~\eqref{Bsmodelloss} yields
\begin{eqnarray}
&&\mathrm{pr}_{n-}(X_\textrm{m};\eta)=\dfrac{\exp\left[-X_\textrm{m}^2\left(\frac{1}{1-\eta}-\frac{B^2}{A}\right)\right]}{n!\pi^{2}\sqrt{(1+2\bar{n})(1-\eta)A}}\sum_{k=0}^{n}\sum_{l=0}^{k}\sum_{p=0}^{k-l}\begin{pmatrix}n\\k\end{pmatrix}\begin{pmatrix}2k\\2l\end{pmatrix}\begin{pmatrix}2(k-l)\\2p\end{pmatrix}
\Gamma\left[n-k+\frac{1}{2}\right]\nonumber\\
&&\qquad\qquad\qquad\Gamma\left[l+\frac{1}{2}\right]\Gamma\left[p+\frac{1}{2}\right]
\left(\dfrac{BX_\textrm{m}}{A}\right)^{2(k-l-p)}\dfrac{(2\bar{n})^{k-l}}{(1+2\bar{n})^{2k-l}}A^{-p},
\end{eqnarray}
where $A=1/(1+2\bar{n})+\eta/(1-\eta)$ and $B=\sqrt{\eta}(1-\eta)$.

An equivalent expression to Eq.~\eqref{Bsmodelloss}, describing the effect of the beamsplitter model is given by the $P$-function transformation: $P'(\beta)=\frac{1}{\eta}P(\beta/\sqrt{\eta})$, where $P'(\beta)$ is the $P$ function of the optical field to be measured. Using \cref{eq:Pfuncnsub}, the $P$ function of $\rho_{n-}$ therefore transforms according to $P'_{n-}(\beta)=P_{n-}(\beta;\bar{n}\rightarrow\eta\bar{n})$, which is a simple rescaling of the initial mean phonon number of the thermal state.  According to Eq.~\eqref{probPfunceq}, the effect of inefficient measurement on the marginal  $\mathrm{pr}_{n-}(X_\textrm{m})$ is also to rescale $\bar{n}$ in the same way: $\bar{n}\rightarrow\eta\bar{n}$. Hence, $\mathrm{pr}_{n-}(X_\textrm{m};\eta)=\mathrm{pr}_{n-}(X_\textrm{m};\bar{n}\rightarrow\eta\bar{n})$.

\subsubsection{Heterodyne detection and the $s$-parameterized Wigner function}

Heterodyne detection projects the optical state $\rho$ entering the detector onto a coherent state $|\alpha\rangle$. An outcome $\alpha\in\mathbb{C}$ occurs with probability proportional to $\Tr(|{\alpha}\rangle\langle\alpha|\rho)=\langle{\alpha}|\rho|{\alpha}\rangle$. Hence, as the Husimi-$Q$ function is defined as $Q(\alpha)=\frac{1}{2\pi}\langle{\alpha}|\rho|{\alpha}\rangle$, heterodyne detection allows one to measure the $Q$ function of the optical state $\rho$. 

In the case of a perfect measurement of the mechanical state, $\eta=1$, heterodyne detection measures the $Q$ function of the mechanical state. When $\eta=1$, the marginals of the $Q$ function are related to the mechanical state marginals $\mathrm{pr}(X_\textrm{m})$ via a convolution with a Gaussian 
\begin{eqnarray}
\mathrm{pr}(X_\textrm{m};s=-1)&=&\int{d}{P}_\textrm{m}~Q(X_\textrm{m},P_\textrm{m})\\
&=&\frac{1}{\sqrt{\pi}}\int{d}{X'}~\mathrm{pr}(X')\mathrm{e}^{{-(X_\textrm{m}-X')^2}}\label{margQ}.
\end{eqnarray}
This equation is again phase-invariant and $s=-1$ refers to condition for the $s$-parameterized Wigner function $W_{s}(X_\textrm{m},P_\textrm{m})$ to equal the $Q$ function.

However, in the case of inefficient detection, $\eta < 1$, the $Q$ function measured by heterodyne detection is smoothed~[U.~Leonhardt and H.~Paul, Phys.~Rev.~A \textbf{48}, 4598 (1993)] according to
\begin{eqnarray}\label{SWigeq}
Q(X_\textrm{m},P_\textrm{m})&=&\frac{1}{\eta}W_{s}(X_\textrm{m}\eta^{-1/2},P_\textrm{m}\eta^{-1/2}),\\
s&=&\frac{1}{\eta}(\eta-2),
\end{eqnarray}
Therefore, inefficient heterodyne measures the $s$-parameterized Wigner function of the mechanical state with $s<-1$.

For a general quantum state, the $s$-parameterized Wigner function can be computed by convolving the $P$ function with a two-dimensional Gaussian  
\begin{align}
    \label{eq:wigners}
    W_{s}(X_\textrm{m},P_\textrm{m}) = \frac{1}{\pi(1-s)}\iint_{-\infty}^{\infty} & {d}{X'_\textrm{m}}{d}{P'_\textrm{m}}~P(X'_\textrm{m},P'_\textrm{m})\nonumber\\
    & \times \mathrm{exp}\left(-\frac{(X_\textrm{m}-X'_\textrm{m})^2+(P_\textrm{m}-P'_\textrm{m})^2}{1-s}\right).
\end{align}
In our case, for $n$-phonon subtraction, the $s$-parametrized Wigner function can be calculated by using \cref{eq:Pfuncnsub} for the $P$ function. The full form of the $s$-parametrized Wigner function for an $n$-phonon subtracted state is then
\begin{align}
    W_s(X_\textrm{m}, P_\textrm{m}) = \frac{2^{-n}\bar{n}^{-1-n}}{\pi^2 (1-s) n!}\iint_{-\infty}^{\infty} & d X_\textrm{m}' d P_\textrm{m}' \bigg[ \left(X_\textrm{m}'^2+P_\textrm{m}'^2\right)^n \nonumber\\
    & \times \mathrm{exp} \left(-\frac{X_\textrm{m}'^2+P_\textrm{m}'^2}{2\bar{n}}-\frac{(X_\textrm{m}'-X_\textrm{m})^2+(P_\textrm{m}'-P_\textrm{m})^2}{1-s}\right)\bigg],
\end{align}
where we have used that $\beta=(X_\textrm{m}+ \mathrm{i}P_\textrm{m})/\sqrt{2}$. Introducing, 
\begin{align}
    G_{s}(X_\textrm{m}, P_\textrm{m}) &= \frac{1}{\pi (1-s)}\mathrm{exp}\left(-\frac{X_\textrm{m}^2+P_\textrm{m}^2}{1-s}\right)
\end{align}
then allows us to write $W_s(X_\textrm{m}, P_\textrm{m})$ more compactly as 
\begin{equation}
    W_s(X_\textrm{m}, P_\textrm{m}) = \left(P_{n-} * G_s\right)(X_\textrm{m}, P_\textrm{m})\ ,
\end{equation}
where $*$ represents the two-dimensional convolution.

\subsubsection{Measured marginal distributions}
Here, we derive an expression that relates the mechanical marginal distribution $\mathrm{pr}(X_\textrm{m})$ to the distribution measured via inefficient heterodyne detection $\mathrm{Pr}(X_\textrm{m})$. Using Eq.~\eqref{SWigeq}, we find the marginal distribution $\mathrm{Pr}(X_\textrm{m})$ of the field measured by heterodyne is
\begin{eqnarray}
\mathrm{Pr}(X_\textrm{m})&=&\int{d}P_\textrm{m}~Q(X_\textrm{m},P_\textrm{m})\\
&=&\frac{1}{\eta}\int{d}P_\textrm{m}~W_{s}(X_\textrm{m}\eta^{-1/2},P_\textrm{m}\eta^{-1/2})\\
&=&\frac{1}{\sqrt{\eta}}\int{d}P~W_{s}(X_\textrm{m}\eta^{-1/2},P)\\
&=&\frac{1}{\sqrt{\pi|s|\eta}}\int{d}{X'}~\mathrm{pr}(X')\mathrm{e}^{-|s|^{-1}(X'-X_\textrm{m}/\sqrt{\eta})^2}\label{measuredPmarg}.
\end{eqnarray}
Where in the last line, we used the relation between the marginal distribution $\mathrm{pr}(X_\textrm{m})$ and the marginal of the $s$-parameterized Wigner function $\mathrm{pr}(X_\textrm{m};s)$:
\begin{eqnarray}
\mathrm{pr}(X_\textrm{m};s)&=&\frac{1}{\sqrt{\pi|s|}}\int{d}{X'}~\mathrm{pr}(X')\mathrm{e}^{{-|s|^{-1}(X_\textrm{m}-X')^2}}\label{Prsconv},
\end{eqnarray}
valid for $s<0$.

As expected, a convolution of $\mathrm{pr}(X_\textrm{m};\eta)$ and the Gaussian in Eq.~\eqref{margQ} yields the same expression for $\mathrm{Pr}(X_\textrm{m})$. This observation allows one to arrive at the neat expression for the measured marginal of the $n$-phonon subtracted state
\begin{eqnarray}
&&\mathrm{Pr}_{n-}(X_\textrm{m})=\frac{1}{\sqrt{\pi}}\int{d}X'~\mathrm{pr}_{n-}(X';\eta)\mathrm{e}^{-(X_\textrm{m}-X')^2}\\
&=&\dfrac{\exp\left(-\frac{X_\textrm{m}^2}{2(1+\eta\bar{n})}\right)}{n!\pi^{2}\sqrt{2(1+\eta\bar{n}})}\sum_{k=0}^{n}\sum_{l=0}^{k}\sum_{r=0}^{k-l}\begin{pmatrix}n\\k\end{pmatrix}\begin{pmatrix}2k\\2l\end{pmatrix}\begin{pmatrix}2(k-l)\\2r\end{pmatrix}
\Gamma\left[n-k+\frac{1}{2}\right]\Gamma\left[l+\frac{1}{2}\right]\Gamma\left[r+\frac{1}{2}\right]\nonumber\\
&&\qquad\qquad\qquad X_\textrm{m}^{2(k-l-r)}\dfrac{(2\eta\bar{n})^{k-l}}{(1+2\eta\bar{n})^{l+r}[2(1+\eta\bar{n})]^{2(k-l)-r}}.
\end{eqnarray}
 Explicitly, the measured marginal distributions of the thermal state $\rho_{\bar{n}}$, single-phonon subtracted state $\rho_{1-}$, and two-phonon subtracted state $\rho_{2-}$ are
\begin{eqnarray}
&&\mathrm{Pr}_{\bar{n}}(X_\textrm{m})=\dfrac{1}{\sqrt{2\pi(1+\eta\bar{n})}}\exp\left[-\frac{X_\textrm{m}^2}{2(1+\eta\bar{n})}\right],\\
&&\mathrm{Pr}_{1-}(X_\textrm{m})=\dfrac{1}{\sqrt{8\pi(1+\eta\bar{n})}}\exp\left[-\frac{X_\textrm{m}^2}{2(1+\eta\bar{n})}\right]\left(\dfrac{2+\eta\bar{n}}{1+\eta\bar{n}}+\dfrac{4\eta\bar{n}}{[2(1+\eta\bar{n})]^2}X_\textrm{m}^2\right),\label{eq:ps1sub}\\
&&\mathrm{Pr}_{2-}(X_\textrm{m})=\dfrac{1}{\sqrt{8\pi(1+\eta\bar{n})}}\exp\left[-\frac{X_\textrm{m}^2}{2(1+\eta\bar{n})}\right]\times\nonumber\\
&&\qquad\qquad\qquad\Bigg(\dfrac{8+8\eta\bar{n}+3(\eta\bar{n})^2}{4(1+\eta\bar{n})^2}+\dfrac{4\eta\bar{n}+(\eta\bar{n})^2}{2(1+\eta\bar{n})^3}X_\textrm{m}^2+\dfrac{(2\eta\bar{n})^2}{[2(1+\eta\bar{n})]^4}X_\textrm{m}^4\Bigg),
\end{eqnarray}
which simplify to the marginals of the mechanical $Q$ functions when $\eta=1$. 

For $\eta\bar{n}\leq2$, a maximum of the measured distribution of the single-phonon subtracted state $\mathrm{Pr}_{1-}(X_\textrm{m})$ occurs at $X_{\mathrm{m}}=0$. While for $\eta\bar{n}>2$, the maxima occur at $X_\textrm{m}=\pm X_{1}$
\begin{eqnarray}
X_{1}=\sqrt{\dfrac{(1+\eta\bar{n})(\eta\bar{n}-2)}{\eta\bar{n}}}.
\end{eqnarray}
and a local minimum is located at $X_{\mathrm{m}}=0$.
Hence, the condition to produce the non-Gaussian ring shape, characterized by a dip at $X_\textrm{m}=0$, is $\eta\bar{n}>2$.

We now sketch how we relate the location of the maxima of $\mathrm{Pr}_{1-}(X_\textrm{m})$ to the radius at which the maxima occur in $W_{s}$. First, we consider the simple expression for the $P$ function of the single-phonon subtracted state by choosing $n=1$ in Eq.~\eqref{eq:Pfuncnsub}. We find that the maxima of $P_{1-}(\beta)$ occur at a radius $\sqrt{2\bar{n}}$ from the origin. Integrating $P_{1-}(\beta)$ over one quadrature then gives an expression for marginal of this $P$ function: $\mathrm{pr}_{1-}(X_\textrm{m};s=+1)$. The maxima of $\mathrm{pr}_{1-}(X_\textrm{m};s=+1)$ occur at $X_{\textrm{m}}=\pm\sqrt{\bar{n}}$. Hence, there is a factor of $\sqrt{2}$ difference between the distance of the maxima from the origin of $\mathrm{pr}_{1-}(X_\textrm{m};s=+1)$ and  $P_{1-}(\beta)$. Second, we then use Eq.~\eqref{Prsconv} and a two-dimensional convolution from  the $P$ function to $W_{s}$, to show that this factor of $\sqrt{2}$ difference persists at the level of $\mathrm{Pr}_{1-}(X_\textrm{m})$ and $W_{s}$. Hence, for $\rho_{1-}$ the maxima of $W_{s}(X_\textrm{m},P_\textrm{m})$ occur at a radius of $r_{1}=\sqrt{2}X_{1}$.

For $\eta\bar{n}>2\sqrt{6}-4$, the maxima for the measured distribution of two-phonon subtracted state $\mathrm{Pr}_{2-}(X_\textrm{m})$ occur at $X_\textrm{m}=\pm X_{2}$
\begin{eqnarray}
X_{2}=\sqrt{\dfrac{1+\eta\bar{n}}{\eta\bar{n}}\left[-4+\eta\bar{n}+\sqrt{2(4+(\eta\bar{n})^2)}\right]},
\end{eqnarray}
and a minimum exists at $X_{\mathrm{m}}=0$.
A similar calculation to the one sketched above gives that for $\rho_{2-}$ the maxima of $W_{s}(X_\textrm{m},P_\textrm{m})$ occur at a radius $r_{2}=\sqrt{2}X_{2}$. In this case, the condition to observe non-Gaussianity in the two-phonon subtracted state is $\eta\bar{n}>2\sqrt{6}-4$. 

In this work we achieve $\eta\bar{n}=4.1$, which satisfies the condition to observe non-Gaussianity in both the single- and two-phonon subtracted state.

\subsection{Temporal evolution of the heterodyne signal variance for two-phonon subtraction}

In order to calculate the temporal evolution of the measured heterodyne signal variance for the case of two-phonon subtraction to a mechanical thermal state, we proceed in a similar manner to the calculation presented in the Supplementary Material of [Enzian \emph{et al.}, Phys.~Rev.~Lett. \textbf{126}, 033601 (2021)] for the case of single-phonon subtraction. 

\begin{figure}[tb!]
    \centering
    \includegraphics[width=80mm]{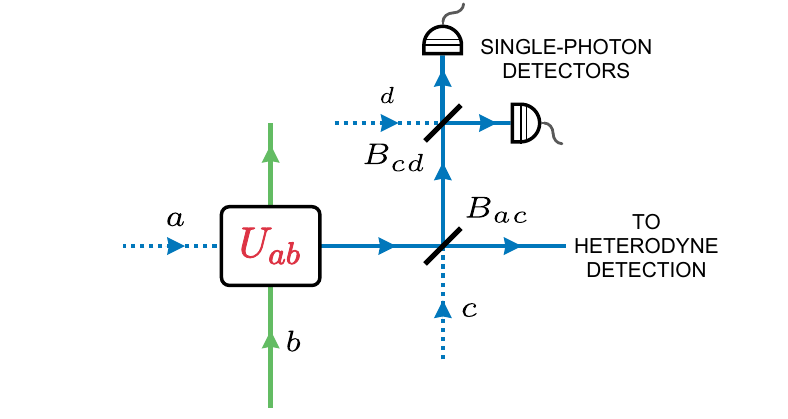}
    \caption{\small 
    Simplified schematic of the experimental setup used to implement and characterize two-phonon subtraction from a mechanical thermal state. Here, $a$ represents the optical cavity mode of the scattered anti-Stokes signal, $b$ is the mechanical mode, and $U_{ab}$ is a unitary corresponding to a light-mechanics beamsplitter-like interaction. An auxiliary mode $c$ has been introduced which participates in the optical beamsplitter, described by the unitary $B_\mathrm{ac}$, dividing the light between the single-photon detectors and the heterodyne measurement.
    } \label{SupFig: Detection model}
\end{figure}

Fig.~\ref{SupFig: Detection model} shows a simplified model of the joint click-dyne detection scheme considered. If the optical fields impinging on the single-photon detectors are weak, such that the probability of more than one photon arriving at a single detector (per gate duration) is negligible, the action of the detectors is well described by projection onto single-photon Fock states. For our scheme, this corresponds the measurement operator
\begin{eqnarray}
\Upsilon 
&=& 
\langle1|_\text{c} \langle1|_\text{d} 
B_\text{cd} 
|0\rangle_\text{d} 
B_\text{ac} 
|0\rangle_\text{c}
\simeq
\langle2|_\text{c} \, B_\mathrm{ac} \, |0\rangle_\text{c}
\end{eqnarray}
where $B_\text{ac}$ and $B_\text{cd}$ are beamsplitter unitaries between modes $a$ and $c$, and $c$ and $d$, respectively.

The mean quadrature variance of the optical cavity mode for a two-phonon subtracted thermal mechanical state is computed using
\begin{eqnarray}
\label{eq:condvar}
\langle X^2_\text{cav}(\tau)\rangle 
= 
\Tr( \rho_\text{c} \, X^2_\text{cav} ) \ ,
\end{eqnarray}
where $\rho_\text{c}$ is the density operator of the conditioned state, given by
\begin{eqnarray}\label{eq:rhoc2subinitial}
\rho_\text{c}
&=&
\frac{1}{\mathcal{P}} 
\langle2|_\text{c}
B_\text{ac} |0\rangle_\text{c} \,
U_\text{ab} \,
\rho_{\bar{n}} \otimes |0\rangle\langle0|_\text{a} \, 
U_\text{ab}^\dagger \,
\langle0|_\text{c} B_\text{ac}^\dagger \,
|2\rangle_c
\ .
\end{eqnarray}
The normalization $\mathcal{P}$ corresponds to the heralding probability, and $\rho_{\bar{n}} \otimes |0\rangle\langle0|_\text{a}$ is the initial state of the whole system, where $\rho_{\bar{n}}$ is the density operator of the thermal mechanical state before the two-fold subtraction event, and $|0\rangle\langle0|_\text{a}$ is the vacuum state of the optical cavity mode.

For a weak signal field arriving at the detectors, we take 
\begin{eqnarray}
B_\mathrm{ac} \simeq 
1 + ir(ac^\dagger~+~ca^\dagger) - r^2(ac^\dagger~+~ca^\dagger)^2/2
\end{eqnarray}
where $r$ is the amplitude reflectivity parameter of the optical beamsplitter, such that $\langle2|_\text{c} \, B_\mathrm{ac} \, |0\rangle_\text{c} = r^2 a^2 / \sqrt{2}$.
Substituting this result into Eq.~\eqref{eq:rhoc2subinitial}, one obtains for the conditioned state
\begin{eqnarray}
\rho_\text{c}
&=&
\frac{1}{\mathcal{P}} 
\left( \frac{r^2}{\sqrt{2}}\right)^2 \, 
a^2 \,
U_\text{ab} \,
\rho_{\bar{n}} \otimes |0\rangle\langle0|_\text{a} \,
U_\text{ab}^\dagger \,
(a^\dag)^2 
\ ,
\end{eqnarray}
with the $\mathcal{P} = r^4 \langle (a^\dag)^2 a^2 \rangle / 2$, such that the quadrature variance of the optical field is given by
\begin{eqnarray}
\label{eq:xcavcond2}
\langle X^2_\text{cav}(\tau)\rangle 
= 
\frac{1}{\mathcal{P}} 
\left( \frac{r^2}{\sqrt{2}}\right)^2
\Tr\big( 
\rho_{\bar{n}} \otimes |0\rangle\langle0| \,\,
U_\text{ab}^\dagger \, (a^\dag)^2  \, 
X^2_\text{cav} \, 
a^2 \, U_\text{ab} 
\big) \ .
\end{eqnarray}

At this point, in order to capture the full cavity dynamics, solutions to the Heisenberg--Langevin equations of motion can be inserted into Eq.~\eqref{eq:xcavcond2} such that
\begin{eqnarray}\label{eq:2subxcav}
\langle X^2_\text{cav} (\tau) \rangle 
&=& 
\frac{ 
\langle 
a^\dag(0) a^\dag(0) \, 
X_\text{cav}^2(\tau) \, 
a(0) a(0) 
\rangle 
}{
\langle a^\dag(0) a^\dagger(0) a(0) a(0) \rangle^2}  \  \ ,
\end{eqnarray}
which, after inserting $X_{\text{cav}} = (a + a^\dag)/\sqrt{2}$, and applying the commutation rules for the creation and annihilation operators of bosonic modes, becomes
\begin{eqnarray}
\langle X^2_\text{cav} (\tau) \rangle 
&=& 
\frac{1}{2} + 
\frac{ \langle a_0^\dag a_0^\dag a_\tau^\dag a_\tau a_0 a_0 \rangle }{\langle a_0^\dag a_0\rangle^2}  \  \ ,
\end{eqnarray}
where we have introduced the notation $a_\tau \equiv a(\tau)$ for brevity, with $\tau=t-t_{0}=0$ denoting the heralding time of a two-phonon subtraction operation.

To proceed, the Isserlis--Wick theorem for Gaussian fields is applied to obtain
\begin{eqnarray}
\langle 
a_0^\dag a_0^\dag 
a_\tau^\dag a_\tau 
a_0 a_0 
\rangle 
&=& 
2 \, \langle a_0^\dag a_0 \rangle^2 \langle a_\tau a_\tau^\dag \rangle 
+ 
4 \langle a_0^\dag a_0 \rangle |\langle a_0^\dag a_\tau \rangle |^2 \ ,
\end{eqnarray}
such that
\begin{eqnarray}
\label{eq:tripling}
\langle X^2_\text{cav}(\tau)\rangle  
&=& 
\frac{1}{2} + 
\langle a_0^\dag a_0 \rangle + \frac{2 |\langle a_0^\dag a_\tau \rangle |^2 }{\langle a_0^\dag a_0\rangle} \ .
\end{eqnarray}

For a light-mechanics beamsplitter-like interaction, with Hamiltonian $H/\hbar = G(ab^\dagger + ba^\dagger)$, the solution to the Heisenberg--Langevin equation of motion for the optical cavity mode is given by Eq.~(21) in the Supplementary Material of [Enzian \emph{et al.}, Phys.~Rev.~Lett. \textbf{126}, 033601 (2021)] as
\begin{align}\label{eq:BSsolution}
        a(t) 
        =
        \sqrt{2 \kappa} \,
        \Big( 
        e^{-\kappa t} \, \Theta(t) 
        \Big) 
        * a_\text{in}(t)
        -  
        \frac{i G \sqrt{2 \gamma}}{\kappa-\gamma} \,
        \bigg(
        \Big( e^{-\gamma t} - e^{-\kappa t} \Big) \, \Theta(t) \bigg) 
        * 
        b_\text{in}(t) \ .
\end{align}
such that
\begin{eqnarray}
\langle a_0^\dag a_\tau \rangle 
&=& 
\frac{\bar{n}_{\mathrm{th}} G^2}{\kappa(\kappa+\gamma)} 
\left( 
\frac{
\kappa e^{-\gamma |\tau|} - \gamma e^{-\kappa|\tau|} 
}{
\kappa-\gamma
}
\right) \ ,
\end{eqnarray}
and
\begin{eqnarray}
\langle a_0^\dag a_0 \rangle 
&=& 
\frac{\bar{n}_{\mathrm{th}} G^2}{\kappa(\kappa+\gamma)} \ .
\end{eqnarray}

Inserting the above correlations back into Eq.~\eqref{eq:tripling}, we obtain for the expected quadrature variance of the optical cavity mode for a two-phonon subtracted thermal mechanical state
\begin{eqnarray}
&&\langle X^2_\text{cav}(\tau) \rangle 
=
\frac{1}{2} + 
\frac{\bar{n}_{\mathrm{th}} G^2}{\kappa(\kappa+\gamma)} \left( 1 + 2\left( \frac{\kappa e^{-\gamma|\tau|} - \gamma e^{-\kappa|\tau|}}{\kappa-\gamma}  \right)^2 \right) 
 \ .
\end{eqnarray}

Next, we note that in the experiment the measured signal is not the intra-cavity field but rather the output mode of the cavity. After accounting for this linear transformation, along with other losses in the system, we normalize such that variance of the optical vacuum is equal to 1, to arrive at 
\begin{eqnarray} \label{eq:twosubvarianceresult}
\sigma_{2-}^2(\tau)
&=&
1+\eta\bar{n}_{\mathrm{th}}\left(1+2\left(\dfrac{\kappa\mathrm{e}^{-\gamma|\tau|}-\gamma\mathrm{e}^{-\kappa|\tau|}}{\kappa-\gamma}\right)^2\right)
\end{eqnarray}
for the measured heterodyne variance of a two-phonon subtracted thermal state, where $\eta$ is the overall measurement efficiency. Similarly, for the case of single-phonon subtraction we also have (see [Enzian \emph{et al.}, Phys.~Rev.~Lett. \textbf{126}, 033601 (2021)]  for further details)
\begin{eqnarray} \label{eq:onesubvarianceresult}
\sigma_{1-}^2(\tau)
&=&
1+\eta\bar{n}_{\mathrm{th}}\left(1+\left(\dfrac{\kappa\mathrm{e}^{-\gamma|\tau|}-\gamma\mathrm{e}^{-\kappa|\tau|}}{\kappa-\gamma}\right)^2\right) \ .
\end{eqnarray}

At the time of the heralding event, the mechanical contribution to the measured heterodyne variance is 
\begin{equation}
\label{eq:variance_doubling_tripling}
\frac{\sigma_{n-}^2(\tau=0) - 1}{\sigma_{n-}^2(\tau\rightarrow\infty) - 1}
= 
1+n \quad 
\end{equation}
for $n \in \set{1,2}$, showing that the effective mean occupation of the mechanical oscillator doubles and triples for the one- and two-subtraction events, respectively. In the long time limit, Eqs.~\eqref{eq:twosubvarianceresult} and \eqref{eq:onesubvarianceresult} both tend towards the measured heterodyne variance of the thermal state $\sigma^2=\eta\bar{n}_\mathrm{th}+1$. The variance of the measured thermal distribution and knowledge of $\bar{n}_\mathrm{th}$ therefore allows one to determine the efficiency $\eta=(\sigma^2-1)/\bar{n}_\mathrm{th}$ and $s$ parameter $s=(\eta-2)/\eta$. With the knowledge of the measurement efficiency, the heterodyne quadrature signal can be normalized to units of mechanical zero-point fluctuations.

\subsection{Experimental details and the BaF$_2$ microresonator}

\subsubsection{Experimental setup}

\begin{figure}[b]
    \centering
    \includegraphics[width=145mm]{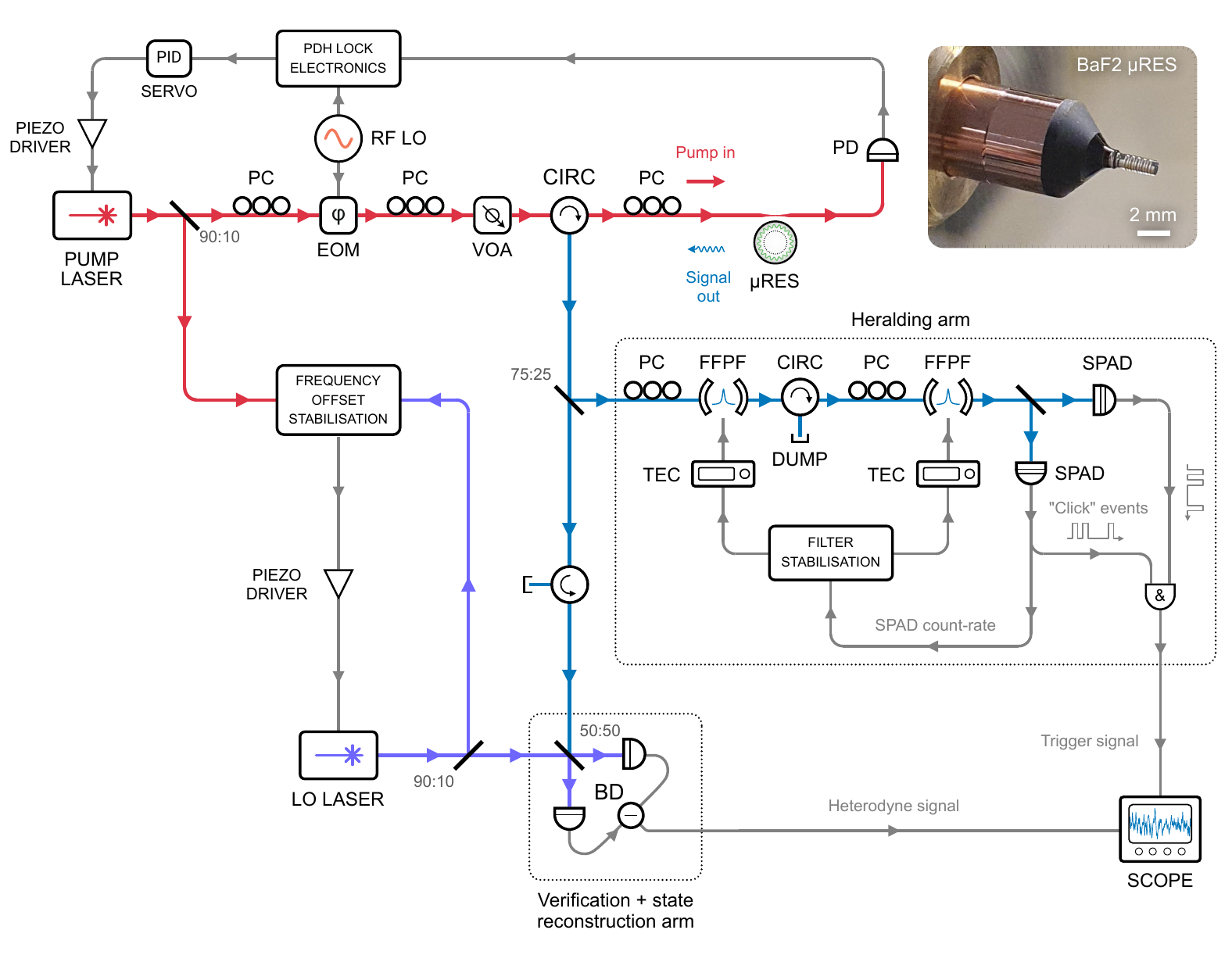}
    \caption{
    Detailed schematic of the fiber-based experimental setup, and (inset) image of BaF$_2$ microresonator. 
    PC: polarisation controller, EOM: electro-optic modulator, VOA: variable optical attenuator, CIRC: circulator, PD: photodiode, $\mu$RES: microresonator, SERVO: servo controller, DUMP: beam dump, FFPF: fiber Fabry-Perot filter, TEC: thermo-electric cooler, LO: local oscillator, BD: balanced photodetector.
    } \label{SupFig:detailed_setup}
\end{figure}

In Figure~\ref{SupFig:detailed_setup} a detailed schematic of the experimental setup and an image of the crystalline microresonator is shown, and in \cref{tab:expparameters} the key experimental parameters are summarized. The pump laser is locked to its respective cavity resonance using the Pound--Drever--Hall laser locking technique. The backscattered Brillouin anti-Stokes light emerging from the cavity is then separated from the pump-field using an optical circulator. This anti-Stokes signal is then split into two arms using a 75:25 beamsplitter: one arm for single- and two-photon detection (heralding arm), and one arm for heterodyne detection (verification/state tomography arm).

In the heralding arm, two single-photon avalanche diodes (SPADs) are used. These detectors are operated using the parameters listed in \cref{tab:expparameters}. In order to suppress spurious pump-photons that can be elastically scattered into the signal mode and make their way to the detectors, two fiber-based Fabry--Perot optical filters are used; both of these filters possess a free spectral range of 25~GHz and an intensity full width at half-maximum of 120~MHz. Using the count rate measured on the SPADs, the filters are continuously tuned to be on-resonance with the signal field using thermo-electric coolers. Furthermore, an optical circulator is placed in between the two filters in order to prevent the formation of an unwanted standing wave cavity due to the back reflections.
 
In the heterodyne detection arm, a second laser is used for the local oscillator. This laser, which is blue-detuned from the anti-Stokes signal field by 200~MHz, is frequency stabilised with respect to the pump-field by continuously monitoring the beat-note between the two lasers and applying a corrective voltage to the fast-piezo tuning port of the laser. The time traces captured on the oscilloscope were \SI{4}{\micro\second} long at a sampling rate of 3.125~GS/s, allowing the temporal dynamics of the heterodyne signal to be resolved at a much faster timescale than the mechanical decay rate.

\begin{table}[tb]
    \centering
        \caption{Experimental parameters.}
        \label{tab:expparameters}
    \begin{tabular}{ll}\hline\hline
    \textbf{Parameter} & \textbf{Value} \\
    \hline
    Sample temperature & 300 K \\
    Input pump power & $\simeq$ 9 mW \\
    Overall measurement efficiency, $\eta$ & 0.91\%\\
    Taper transmission & 0.89 \\
    Filtering arm transmission  & 0.15 \\
    Heterodyne arm overall detection efficiency \ & 0.365 \\
    FFPF linewidth (FWHM) &  120 MHz \\
    FFPF free spectral range &  25 GHz \\
    SPAD quantum efficiency  & 0.125 \\
    SPAD gate rate & 50 kHz \\
    SPAD gate length & 3.5 ns \\
    SPAD dead time & \SI{18}{\micro\second}\\
    SPAD dark count rate  & $\simeq 1$ s$^{-1}$ \\
    Count rate  & $260(60)$ s$^{-1}$ \\
    Coincidence rate & $\simeq 2$ s$^{-1}$ \\
    Heterodyne frequency, $\omega_\text{het}/2\pi$ & 214 MHz \\
    Balanced detector bandwidth &  $\simeq$ 400 MHz \\
    Recorded time trace length &  \SI{4}{\micro\second} \\
    Sampling rate &  \SI{3.125}~GS s$^{-1}$ \\
    Number of time traces per operation \,\, & \num{2.4e5} \\
    \hline\hline
    \end{tabular} \\
\end{table} \,

\subsubsection{BaF$_2$ microresonator}

The barium fluoride microresonator used for this experiment was fabricated using a diamond nano-lathe and is shown in Fig.~\ref{SupFig:detailed_setup}. Starting from a 3~mm BaF$_2$ cylinder (glued to a copper mounting cylinder), the resonator is first machined to the required resonator diameter before precise cutting of the resonator mode volumes (``bulges''). Post-cutting, the resonator is spun in the lathe and hand polished using a cleaning slurry followed by isopropanol. The dimensions of the resonator are given in \cref{tab:sysparameters}.

This crystalline material is chosen owing to the high optical quality factor whispering gallery modes achievable in these systems [Lin \emph{et al.}, Opt.~Lett. \textbf{39}, 6009--6012 (2014)] and the low acoustic damping rates available in crystalline materials compared to amorphous materials [Ohno \emph{et al.} Rev.~Sci.~Instrum. \textbf{12}, 123104 (2006); Galliou \emph{et al.}, Sci.~Rep. \textbf{3}, 2132 (2013); Renninger \emph{et al.}, Nat.~Phys. \textbf{14}, 601 (2018)].

\subsection{System characterization and parameters}\label{supp:section: antistokes gamma measurement}
To characterize our optomechanical system, we measure the spectrum of the thermally scattered anti-Stokes signal at a range of input powers. In the limit of weak coupling, it can be shown that the power spectral density of the scattered signal is given by
\begin{eqnarray}
S_{X\!X}(\omega)
&=& 
\int_{-\infty}^{\infty} \! d\omega \, 
\langle \widetilde{X}^\dagger(\omega) \widetilde{X} (\omega) \rangle
\nonumber \\
&\propto&
|\chi_\mathrm{bb}(\omega-\omega_\mathrm{H})|^2 + 
|\chi_\mathrm{bb}(-\omega-\omega_\mathrm{H})|^2
\ , \
\end{eqnarray}
where $\widetilde{X}$ is the Fourier transform of the $X$-quadrature, $\omega_\mathrm{H}$ is the heterodyne frequency, and
\begin{eqnarray}
\chi_\mathrm{bb}(\omega)
&=&
\frac{\sqrt{2\gamma}}{i\omega + \gamma_\mathrm{eff}}
\end{eqnarray}
is the mechanical susceptibility in the limit of $\gamma \ll \kappa_2$, where $\kappa_2$ is the anti-Stokes resonance decay rate. The effective mechanical amplitude decay rate is given by 
\begin{equation}
    \label{eq:gamma-eff}
    \gamma_\mathrm{eff} 
    = \gamma \left(1+\frac{G^2}{\kappa_2\gamma}\right) 
    = \gamma (1+C)\ ,
\end{equation}
where $C$ denotes the optomechanical cooperativity, and the coupling rate $G$ is related to the number of intra-cavity photons $N_\text{cav}$ and the single-photon coupling rate $g_0$ by
\begin{equation}
    \label{eq:optomech-coupling}
    G = g_0\sqrt{N_\text{cav}}\ .
\end{equation}

In our experiment, we operate well within the weak-coupling regime and note that the intrinsic mechanical decay rate is much less than the decay rate of the ``scattered into" anti-Stokes optical mode ($\gamma/\kappa_2 \simeq 0.07$). The spectrum of the backscattered anti-Stokes signal is therefore well approximated by a Lorentzian function with a full-width half-maximum given by $2\gamma_\mathrm{eff}$ that scales linearly with the number of intra-cavity pump photons.

\begin{figure}[t]
    \centering
    \includegraphics[width=140mm]{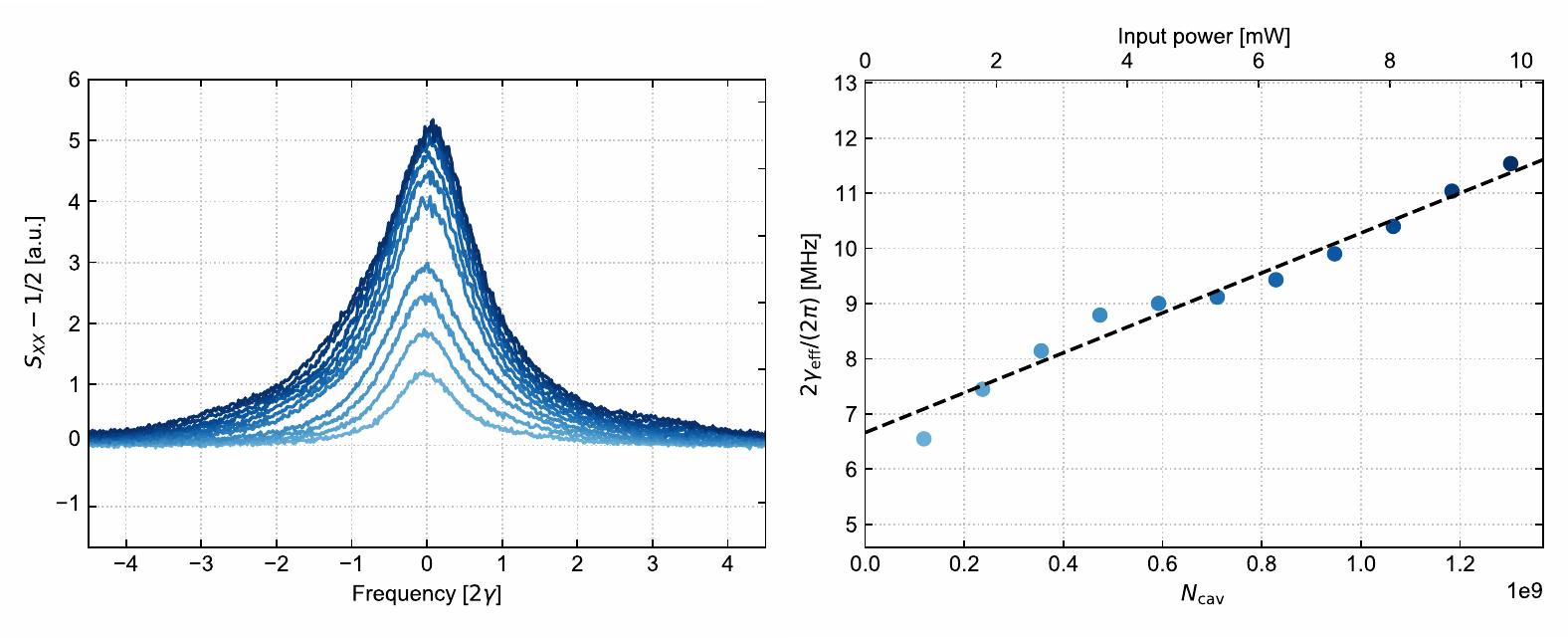}
    \caption{\small 
    (a) Mean spectrum of the thermally scattered anti-Stokes light at various input powers with frequency plotted in units of the intrinsic mechanical decay rate $\gamma$. Darker shades correspond to higher input powers up to a maximum of $10$~mW.
    (b) Effective mechanical linewidth as a function of intra-cavity pump photon number with linear fit indicated by a dashed line.
    } \label{SupFig: Anti-Stokes Spectra}
\end{figure}

The experimental setup used to measure the mechanical linewidth as a function of input power is the same as in Fig.~\ref{Fig:SchemeSetup} of the main text. To obtain the spectra of the scattered signal we Fourier-transform time traces from the output of the balanced heterodyne detector at each input power. To observe optomechanical broadening of the signal linewidth, the measurement is performed for input powers up to $\simeq 10$~mW, as shown in Fig.~\ref{SupFig: Anti-Stokes Spectra}. Fitting to $\gamma_\mathrm{eff}$, we obtain an intrinsic mechanical decay rate of $2\gamma/2\pi = 6.52(80)$~MHz, which is in good agreement with room-temperature Brillouin linewidths reported in similar materials [T.~Sonehara \emph{et al.}, J.~Opt.~Soc.~Am.~B \textbf{24}, 1193--1198 (2007)]. Furthermore, from the gradient, using \cref{eq:gamma-eff,eq:optomech-coupling}, we also obtain $g_0/2\pi = 296(37)$~Hz.

The linewidths of the optical cavity modes are determined by sweeping the pump laser in frequency at low power. The pump and anti-Stokes optical resonances are measured to be $2\kappa_1/2\pi = 14.1$~MHz and $2\kappa_2/2\pi = 93.7$~MHz, respectively. Note that in order to compute the pump intra-cavity photon number we use $N_\mathrm{cav}\!~=~\!\eta_\mathrm{c,1} P_\mathrm{in}/(\kappa_1\hbar\omega)$, where $\kappa_1$ is the (amplitude) decay rate of the optical pump mode and $\eta_\mathrm{c,1} = 2\kappa_1^\mathrm{e}/\kappa_1$. The extrinsic optical decay rate is calculated using $\kappa_1^\mathrm{e} = \kappa_1 (1-\sqrt{T_0})/2$ where $T_0$ is the on-resonance transmission of the pump-mode (for under-coupled coupling conditions, as pertinent to this work). 

Finally, we quantify the amount of optomechanical sideband-cooling that has occurred prior to the single and two-phonon subtraction events. In the limit of weak coupling, the effective mechanical occupancy in the steady-state is given by
\begin{eqnarray}\label{eq:coolingfactor}
\bar{n} 
&=&
\frac{ \bar{n}_\text{th} }{ 2\pi}
\int_{-\infty}^{\infty} \!\! d\omega \,\,
|\chi_\mathrm{bb}(\omega)|^2
=
\frac{\bar{n}_\text{th}}{1+C},
\end{eqnarray}
where $\bar{n}_\text{th}$ is the initial occupation of the mechanical mode in the absence of any optomechanical coupling and is given by $\bar{n}_\text{th} = k_\text{b} T / (\hbar \omega_\text{m})$ in the limit of high temperature $T$ ($k_\text{b} T \gg \hbar \omega_\text{m}$), where $k_\mathrm{b}$ is the Boltzmann constant and $\omega_\mathrm{m}$ is the angular frequency of the mechanical mode. For the $8.16$~GHz acoustic wave used in this work we have $\bar{n}_\text{th} \simeq 766$. For $G/2\pi \simeq 10$~MHz, as described in the main text, the resulting steady-state occupancy of the unconditioned mechanical state is given by $\bar{n} = 453(52)$, corresponding to a sideband-cooling factor of $\simeq$ 0.6.

A summary of the system parameters can be found in \cref{tab:sysparameters}.

\begin{table}[htb]
    \centering
        \caption{Summary of system parameters.}
    \label{tab:sysparameters}
    \begin{tabular}{lll}\hline\hline
         \textbf{Parameter} &\textbf{Symbol} &  \textbf{Value} \\
         \hline
         Resonator major diameter & $D_\text{res}$ & 0.936 mm \\
         Resonator minor radius & $r_\text{res}$ & 40 $\mu$m \\
         Pump wavelength & $\lambda_\text{p}$ & 1550 nm \\
         Mechanical frequency & $\omega_\text{m}/2\pi$ & $8.16$ GHz \\
         Mechanical linewidth & $2\gamma/2\pi$ & $6.52(80)$ MHz \\
         Mechanical $Qf$ product & $Qf$ & \num{1.02e13} Hz\\
         Pump mode linewidth & $2\kappa_1/2\pi$ & 14.1 MHz \\ 
         Pump mode external coupling & 2$\kappa_1^\text{e}/2\pi$ & 5.3 MHz \\ 
         Signal (aS) mode linewidth & $2\kappa_2/2\pi$ & 93.7 MHz \\ 
         Signal (aS) mode external coupling \hspace{4pt} & 2$\kappa_2^\text{e}/2\pi$ & 11.7 MHz \\ 
         Taper coupling efficiency (aS mode)  \hspace{4pt} & $\eta_\text{c} = 2\kappa_2^\text{e}/\kappa_2$ \hspace{4pt} & $\simeq$ 0.25 \\
         Optomechanical coupling rate& $g_0/2\pi$ & $296(37)$ Hz \\ 
         Intra-cavity pump photon number & $N_\text{cav}$ & $\simeq 1.2 \times 10^9$ \\
         Pump-enhanced coupling rate& $G/2\pi$ & $10.3(13)$ MHz \\  
         Optomechanical cooperativity & $C$ & $0.69(19)$ \\ 
         Mean initial phonon number & $ \bar{n}_\text{th}$ & 766 \\
         Mean effective phonon number & $\bar{n} \rightarrow 2 \bar{n}\,,\,3\bar{n}$ \hspace{4pt} & $453(52)$ $\rightarrow$ 906\,,\,1359 \, \\
         \hline\hline
    \end{tabular}
\end{table} \,

\subsection{Fidelity of single- and multi-phonon subtraction event heralding}

The single-photon detectors used in this work are not photon-number resolving, so as to ensure that the fidelity of the heralded operations is high, the probability of more than one photon arriving at a single detector during a given detector gate-window must be kept low. In this regime, the probability of a single detector registering multi-photon events is extremely small -- and can thus be neglected -- such that any recorded event faithfully corresponds to the detection of a single-photon and only a single-photon. The fidelity of the heralded phonon-subtraction operations is then only affected by the ratio of real-to-dark counts, which for our experiment is $<1$\% (see \cref{tab:expparameters} for details), and the presence of spurious pump-photons which are sufficiently suppressed by the high-finesse optical filters. To add photon-number resolving capabilities to the measurement, coincidence ``clicks'' between multiple detectors are used, allowing multi-photon events to be recorded.

We now compute the mean photon-number in a given detection window for a single-SPAD to quantify the approximations outlined above. The mean photon flux emerging from the cavity is given by
\begin{eqnarray}
\langle F_\mathrm{cav} \rangle 
&=&
2\kappa_2^\mathrm{e} \, \frac{\bar{n}_\mathrm{th} G^2}{\kappa_2(\kappa_2 + \gamma)}\ ,
\end{eqnarray}
which, for the system parameters outlined in the previous section, yields $\langle F_\mathrm{cav} \rangle \sim 10^8 $~s$^{-1}$.

Accounting for the various efficiencies and losses in the system (cf.~\cref{SupFig:detailed_setup}), the rate at which photons arrive at a single-photon detector is given by
\begin{eqnarray}
\langle R_\mathrm{det} \rangle
&=&
\eta_\mathrm{loss}
\eta_\mathrm{bs1}
\eta_\mathrm{filters}
\eta_\mathrm{bs2}
\,
\langle F_\mathrm{cav} \rangle \sim 10^7 \ \si{\per\second}\,
\end{eqnarray}
when inserting the parameters: $\eta_\mathrm{loss} = 0.67$ accounting for the losses in the system up to the 75:25 beamsplitter dividing the signal between the heralding arm and the heterodyne detection arm; $\eta_\mathrm{bs1}=0.25$, the efficiency at which light is split-off off into the heralding arm; $\eta_\mathrm{filters} = 0.15$ accounting for the transmission through the optical filters; and $\eta_\mathrm{bs2} = 0.5$ accounting for the splitting of the light between the two SPADs.

Finally, the mean number of photo-counts per gate at the detector is given by
\begin{eqnarray}
\langle N_\mathrm{det} \rangle
&\approx&
\eta_\mathrm{det}
\langle R_\mathrm{det} \rangle
T_\mathrm{det}\ ,
\end{eqnarray}
where $\eta_\mathrm{det} = 0.125$ and $T_\mathrm{det} = 3.5$~ns are the operating efficiency and gate length of the detector, respectively. Substituting in these values, we find that the mean number of photocounts is 
\begin{equation}
    \langle N_\mathrm{det} \rangle \sim 10^{-2}\ .
\end{equation}
As $\langle N_\mathrm{det} \rangle \ll 1$, the assumption of single-photon events at a single detector is safely valid.

Similarly, it is also useful to note that in our experiment, after accounting for experimentally determined optical circuit and detector efficiencies in the heterodyne detection arm, less than one photon on average is being used for state tomography in the heterodyne measurement within the timescale that the mechanical oscillator is taken out of equilibrium ($\sim$31~ns). It is important to note, that heterodyne detection in itself does not result in phonon subtraction, nor is it used for heralding, and hence does not change the mechanical state.


\begin{thebibliography}{99}

\bibitem{Galliou2013} S. Galliou, M. Goryachev, R. Bourquin, P. Abbe, J. P. Aubry, and M. E. Tobar, Sci. Rep. \textbf{3}, 2132 (2013).

\bibitem{Renninger2018} W. H. Renninger, P. Kharel, R. O. Behunin, and P. T. Rakich, Nat. Phys. \textbf{14}, 601 (2018).

\bibitem{MacCabe2020} G. S. MacCabe \emph{et al.}, Science \textbf{370}, 840 (2020).

\bibitem{Higginbotham2018} A. P. Higginbotham \emph{et al.}, Nat. Phys. \textbf{14}, 1038 (2018).

\bibitem{Mirhosseini2020} M. Mirhosseini \emph{et al.}, Nature \textbf{588}, 599 (2020).

\bibitem{Kim2016} P. H. Kim, B. D. Hauer, C. Doolin, F. Souris, and J. P. Davis, Nat. Commun. \textbf{7}, 13165 (2016).

\bibitem{Monteiro2017} F. Monteiro, S. Ghosh, A. G. Fine, and D. C. Moore, Phys. Rev. A \textbf{96}, 063841 (2017).

\bibitem{Carney2021} D. Carney, A. Hook, Z. Liu, J. M. Taylor, and Y. Zhao, New J. Phys. \textbf{23}, 023041 (2021).

\bibitem{BJK99} S. Bose, K. Jacobs, and P. L. Knight, Phys. Rev. A \textbf{59}, 3204 (1999).

\bibitem{Marshall2003} W. Marshall, C. Simon, R. Penrose, and D. Bouwmeester, Phys. Rev. Lett. \textbf{91}, 130401 (2003).

\bibitem{Bassi2013} A. Bassi, K. Lochan, S. Satin, T. P. Singh, and H. Ulbricht, Rev. Mod. Phys. \textbf{85}, 471 (2013).

\bibitem{Pikovski2012} I. Pikovski \emph{et al.}, Nat. Phys. \textbf{8}, 393 (2012).

\bibitem{Bose2017} S. Bose \emph{et al.}, Phys. Rev. Lett. \textbf{119}, 240401 (2017).

\bibitem{Marletto2017} C. Marletto and V. Vedral, Phys. Rev. Lett. \textbf{119}, 240402 (2017).

\bibitem{Leibfried1996} D. Leibfried \emph{et al.}, Phys. Rev. Lett. \textbf{77}, 4281 (1996).

\bibitem{Fluhmann2019} C. Fl\"{u}hmann \emph{et al.}, Nature \textbf{566}, 513 (2019).

\bibitem{Lvovsky2001} A. I. Lvovsky \emph{et al.}, Phys. Rev. Lett. \textbf{87}, 050402 (2001).

\bibitem{Ourjoumtsev2006} A. Ourjoumtsev, R. Tualle-Brouri, J. Laurat, and P. Grangier, Science \textbf{312}, 83 (2006).

\bibitem{Neergaard2006} J. S. Neergaard-Nielsen, B. Melholt Nielsen, C. Hettich, K. Molmer, and E. S. Polzik, Phys. Rev. Lett. \textbf{97}, 083604 (2006).

\bibitem{Gerrits2010} T. Gerrits \emph{et al.}, Phys. Rev. A \textbf{82}, 031802(R) (2010).

\bibitem{Zavatta2007} A. Zavatta, V. Parigi, and M. Bellini, Phys. Rev. A \textbf{75}, 052106 (2007).

\bibitem{Parigi2007} V. Parigi, A. Zavatta, M. S. Kim, and M. Bellini, Science \textbf{317}, 1890 (2007).

\bibitem{Bogdanov2017} Yu. I. Bogdanov \emph{et al.}, Phys. Rev. A \textbf{96}, 063803 (2017).

\bibitem{Deleglise2008} S. Deleglise \emph{et al.}, Nature \textbf{455}, 510 (2008).

\bibitem{McConnell2015} R. McConnell, H. Zhang, J. Hu, S. Cuk, and V. Vuletic, Nature \textbf{519}, 439 (2015).

\bibitem{Hofheinz2009} M. Hofheinz \emph{et al.}, Nature \textbf{459}, 546 (2009).

\bibitem{Chu2018} Y. Chu, P. Kharel, T. Yoon, L. Frunzio, P. T. Rakich, and R. J. Schoelkopf, Nature \textbf{563}, 666 (2018).

\bibitem{Satzinger2018} K. J. Satzinger \emph{et al.}, Nature \textbf{563}, 661 (2018).

\bibitem{Lee2012} K. C. Lee \emph{et al.}, Nat. Photonics \textbf{6}, 41 (2012).

\bibitem{Fisher2017} K. A. G. Fisher \emph{et al.}, Phys. Rev. A \textbf{96}, 012324 (2017).

\bibitem{Velez2019} S. T. Velez, \emph{et al.}, Phys. Rev. X \textbf{9}, 041007 (2019).

\bibitem{Riedinger2016} R. Riedinger \emph{et al.}, Nature \textbf{530}, 313 (2016).

\bibitem{Cohen2015} J. D. Cohen \emph{et al.}, Nature \textbf{520}, 522 (2015).

\bibitem{Galinskiy2020} I. Galinskiy, Y. Tsaturyan, M. Parniak, E. S. Polzik, Optica \textbf{7}, 718 (2020).

\bibitem{Ringbauer2018} M. Ringbauer \emph{et al.}, New J. Phys. \textbf{20}, 053042 (2018).

\bibitem{Enzian2021} G. Enzian \emph{et al.}, Phys. Rev. Lett. \textbf{126}, 033601 (2021).

\bibitem{Vanner2013} M. R. Vanner, J. Hofer, G. D. Cole, and M. Aspelmeyer, Nat. Commun. \textbf{4}, 2295 (2013).
 
\bibitem{Muhonen2019} J. T. Muhonen, G. R. La Gala, R. Leijssen, and E. Verhagen, Phys. Rev. Lett. \textbf{123}, 113601 (2019).

\bibitem{Suchoi2015} O. Suchoi, K. Shlomi, L. Ella, and E. Buks, Phys. Rev. A \textbf{91}, 043829 (2015).

\bibitem{Rashid2017} M. Rashid, M. Toros, and H. Ulbricht, Quantum Meas. Quantum Metrol. \textbf{4}, 17 (2017).

\bibitem{Vanner2015} M. R. Vanner, I. Pikovski, and M. S. Kim, Ann. Phys. (Berl.) \textbf{527}, 15 (2015).

\bibitem{VannerKim2013} M. R. Vanner, M. Aspelmeyer, and M. S. Kim, Phys. Rev. Lett. \textbf{110}, 010504 (2013).

\bibitem{Barnett2018} S. M. Barnett, G. Ferenczi, C. R. Gilson, and F. C. Speirits, Phys. Rev.  A \textbf{98}, 013809 (2018).

\bibitem{Lin2014} G. Lin, S. Diallo, K. Saleh, R. Martineghi, J.~C. Beugnot, T. Sylvestre, and Y.~K. Chembo, Appl. Phys. Lett. \textbf{105}, 231103 (2014).

\bibitem{Leonhardt1993} U. Leonhardt and H. Paul, Phys. Rev. A \textbf{48}, 4598 (1993); U. Leonhardt, `Measuring the Quantum State of Light' Cambridge University Press (1997).

\bibitem{Enzian2019} G. Enzian \emph{et al.}, Optica \textbf{6}, 7 (2019).

\bibitem{Milburn2016} T. J. Milburn, M. S. Kim, and M. R. Vanner, Phys. Rev. A \textbf{93}, 053818 (2016).

\bibitem{Shomroni2020} I. Shomroni, L. Qiu, and T. J. Kippenberg,
Phys. Rev. A \textbf{101}, 033812 (2020).

\bibitem{Zhan2020} H. Zhan, G. Li, and H. Tan, Phys. Rev. A \textbf{101}, 063834 (2020).

\bibitem{Supp} See the Supplementary Material online for further details.

\bibitem{Ohno2006} S. Ohno, T. Sonehara, E. Tatsu, A. Koreeda, and S. Saikan, Rev. Sci. Instrum. \textbf{12}, {123104} (2006).

\bibitem{Patel2021} R. N. Patel \emph{et al.}, arXiv:2102.04017 (2021).

\end{thebibliography}
\end{document}